\date{May 2023}

\documentclass[twocolumn,aps,prb,longbibliography,amsmath,amssymb,superscriptaddress]{revtex4-2}

\usepackage{graphicx}
\usepackage{dcolumn}
\usepackage{bm}
\usepackage{float}
\usepackage{hyphenat}
\usepackage{hyperref}
\usepackage[dvipsnames]{xcolor}

\newcommand{\units}[1]
{\text{#1}}

\newcommand{\heriotwatt}{Institute of Photonics and Quantum Sciences, SUPA, Heriot-Watt University, Edinburgh EH14 4AS, UK}
\newcommand{\Boston}{Department of Physics, Boston College, Chestnut Hill, MA 02467, USA
}
\newcommand{\oak}{Materials Science and Technology Division, Oak Ridge National Laboratory, Oak Ridge, Tennessee 37831, USA}
\newcommand{\equalcontrib}{These authors contributed equally to this work}
\newcommand{\corresp}{Corresponding author: M.Brotons\_i\_Gisbert@hw.ac.uk}

\begin{document}

\title{Optical contrast analysis of $\alpha$-RuCl$_3$ nanoflakes on oxidized silicon wafers}

    \author{Tatyana V. Ivanova}
    \thanks{\equalcontrib}
    \affiliation{\heriotwatt}
    \author{Daniel Andres-Penares}
    \thanks{\equalcontrib}
    \affiliation{\heriotwatt}
    \author{Yiping Wang}
    \affiliation{\Boston}
    \author{Jiaqiang Yan}
    \affiliation{\oak}
    \author{Daniel Forbes}
    \affiliation{\heriotwatt}
    \author{Servet Ozdemir}
    \affiliation{\heriotwatt}
    \author{Kenneth S. Burch}
    \affiliation{\Boston}
    \author{Brian D. Gerardot}
    \affiliation{\heriotwatt}
    \author{Mauro Brotons-Gisbert}
    \thanks{\corresp}
    \affiliation{\heriotwatt}
    

\begin{abstract}
$\alpha$-RuCl$_3$, a narrow-band Mott insulator with large work function, offers intriguing potential as a quantum material or as a charge acceptor for electrical contacts in van der Waals devices. In this work, we perform a systematic study of the optical reflection contrast of $\alpha$-RuCl$_3$ nanoflakes on oxidized silicon wafers and estimate the accuracy of this imaging technique to assess the crystal thickness. Via spectroscopic micro-ellipsometry measurements, we characterize the wavelength-dependent complex refractive index of $\alpha$-RuCl$_3$ nanoflakes of varying thickness in the visible and near-infrared. Building on these results, we simulate the optical contrast of $\alpha$-RuCl$_3$ nanoflakes with thicknesses below $100$ nm on SiO$_2$/Si substrates under different illumination conditions. We compare the simulated optical contrast with experimental values extracted from optical microscopy images and obtain good agreement. Finally, we show that optical contrast imaging allows us to retrieve the thickness of the RuCl$_3$ nanoflakes exfoliated on an oxidized silicon substrate with a mean deviation of $-0.2$ nm for thicknesses below 100 nm with a standard deviation of only 1 nm. Our results demonstrate that optical contrast can be used as a non-invasive, fast, and reliable technique to estimate the $\alpha$-RuCl$_3$ thickness.
\end{abstract}

\maketitle

\section{\label{sec:level1}Introduction}

Van der Waals materials offer an unprecedented possibility to design and fabricate two-dimensional (2D) heterostructure devices due to a broad palette of atomically thin crystals with unique optical, electric, and magnetic properties. Among the different van der Waals layered crystals, the narrow-band Mott insulator $\alpha$-RuCl$_3$ has emerged as a new building block for achieving local charge control in 2D van der Waals heterostructures. Its narrow electronic bands and large work function make $\alpha$-RuCl$_3$ (hereafter RuCl$_3$) an excellent atomic crystalline charge acceptor, enabling modulation doping of several atomically thin crystals even when just a single RuCl$_3$ layer is employed \cite{wang2020modulation}. Recently, RuCl$_3$ has been used to demonstrate low-resistance Ohmic contact for p-type WSe$_2$ at low temperature and low carrier densities \cite{pack2023charge, xie2023low}, overcoming the well-known challenges associated with achieving low-resistance electrical contacts to monolayer transition-metal dichalcogenides semiconductors \cite{allain2015electrical,wang2022making}. Furthermore, the unique electronic structure of RuCl$_3$ itself has attracted a lot of attention due its complex magnetic interactions, including access to a Kitaev quantum spin liquid phase \cite{banerjee2016proximate,banerjee2017neutron,do2017majorana,wang2020range}. Combined, these properties position 2D RuCl$_3$ as a versatile building block in the van der Waals platform, with the potential to incorporate novel magnetic states \cite{lee2006modulation} and the resulting topological excitations into 2D devices \cite{wang2020modulation}. 

Similar to other 2D materials,  high quality RuCl$_3$ nanoflakes are usually obtained by mechanical exfoliation from bulk crystals on oxidized silicon wafers \cite{Novoselov2005gr}, which typically results in stochastic nanoflakes with randomly varying thicknesses, lateral dimensions, and spatial locations on the substrate. In this context, an experimental technique that allows a fast, non-destructive, and accurate estimation of the thickness of the RuCl$_3$ nanoflakes would represent a valuable asset in the fabrication of 2D heterostructures incorporating this material. Although optical microscopy is routinely employed as initial step in the thickness estimation of mechanically exfoliated RuCl$_3$ \cite{yang2023magnetic,xie2023low}, to our knowledge, a systematic study of the optical reflection contrast of RuCl$_3$ nanoflakes on oxidized silicon wafers is still missing.

Here, we perform spectroscopic ellipsometry measurements at room temperature on mechanically exfoliated RuCl$_3$ nanoflakes of different thickness for photon wavelengths in the visible and near infrared range. Our measurements allow us to estimate the wavelength-dependent complex refractive index of RuCl$_3$ nanoflakes in this spectral range, confirming their thickness-independent optical properties. Building on these results, we exploit the estimated complex refractive index to calculate the optical contrast of RuCl$_3$ nanoflakes between $\sim2-100$ nm on SiO$_2$/Si substrates under different illumination conditions. We compare the calculated optical contrast with the thickness-dependent experimental optical contrast extracted from optical microscopy images, showing good agreement between the simulated and experimental values. The good agreement of the calculated and experimental optical contrast allows us to predict the monochromatic illumination wavelengths that maximise the optical contrast response of RuCl$_3$ nanoflakes below 10 nm for two standard thicknesses (90 and 295 nm) of the SiO$_2$ layer on a Si substrate. Finally, we show that for RuCl$_3$ nanoflakes in the range $2-100$ nm imaged with our calibrated standard optical microscope, the crystal thickness estimated exclusively from the optical contrast analysis shows only a mean deviation of $-0.2$ nm from the value measured by a combination of atomic force microscopy (AFM) and spectroscopic ellipsometry, with a standard deviation close to the thickness of a single RuCl$_3$ monolayer. Our results provide valuable information about the optical properties of 2D RuCl$_3$ flakes in the visible and near infrared, which are crucial to exploit this material in 2D nanodevices. Moreover, we show that optical contrast can be used as a non-invasive, fast, and reliable technique to estimate RuCl$_3$ crystal thicknesses.

\section{\label{sec:level1}Complex refractive index of R\lowercase{u}C\lowercase{l}$_3$ in the visible and near-infrared}

\begin{figure*}
\includegraphics[scale=0.56]{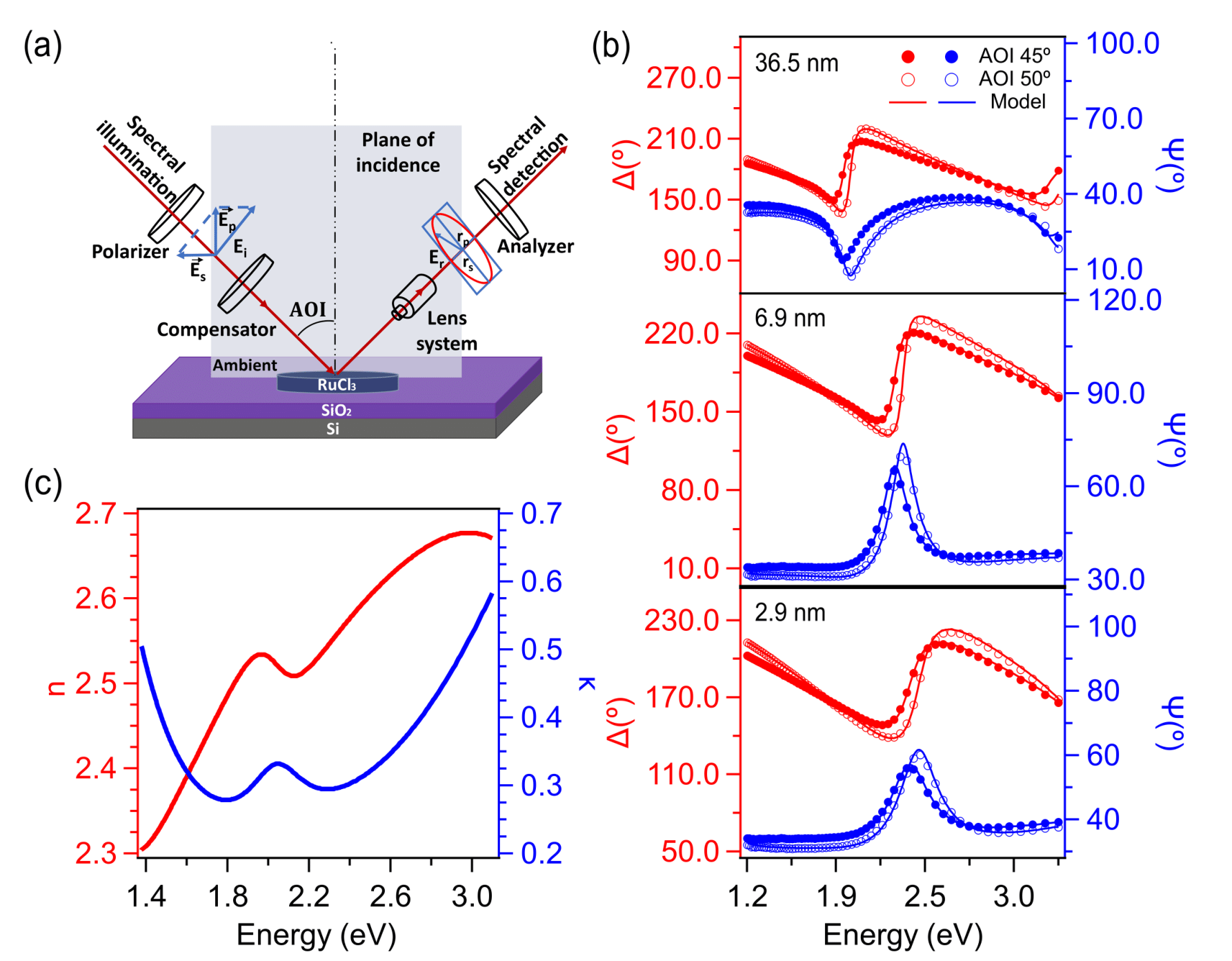}
\caption{\label{fig1}(a) Schematic representation of the experimental spectroscopic ellipsometry setup. (b) $\Delta$ (red) and $\Psi$ (blue) ellipsometric angles measured under AOIs of 45$^\circ$ (filled circles) and 50$^\circ$ (empty circles) for RuCl$_3$ nanoflakes with thicknesses of 2.9 (bottom), 6.9 (middle) and 36.5 nm (top). The solid lines represent the fit of the experimental data using a Mueller matrix formalism. (c) Estimated complex refractive index of RuCl$_3$ obtained from a simultaneous fitting of the $\Delta$ and $\Psi$ angles shown in (b) for different crystal thicknesses.}
\end{figure*}

We begin by performing spectroscopic micro-ellipsometry measurements at room temperature on RuCl$_3$ nanoflakes with thicknesses between $\sim$3 and 40 nm for photon wavelengths in the visible and near infrared ($\sim$1.4 - 3.1 eV). The RuCl$_3$ nanoflakes were obtained by mechanical exfoliation from bulk crystals directly on Si substrates with a top SiO$_2$ layer with a nominal thickness of 295 nm. The thickness of each nanoflake was measured by AFM. The spectroscopic micro-ellipsometry measurements were carried out using an Accurion EP4 imaging ellipsometer with a spatial resolution of $\sim1$ $\mu m$. Figure \ref{fig1}(a) shows a sketch of the experimental measurement setup. The ellipsometric $\Delta$ and $\Psi$ angles of the multilayer system (RuCl$_3$/SiO$_2$/Si) were measured as a function of the energy and the angle of incidence (AOI) of the illumination light. Figure \ref{fig1}(b) shows the $\Delta$ (red) and $\Psi$ (blue) angles measured under AOIs of 45$^\circ$ (filled circles) and 50$^\circ$ (empty circles) for RuCl$_3$ nanoflakes with thicknesses of 2.9, 6.9, and 36.5 nm. We note that these measurements were repeated for different in-plane rotations of the RuCl$_3$ samples, leading to imperceptible differences on the measured $\Delta$ and $\Psi$ angles, as expected from the in-plane isotropic optical response of the material.

For a quantitative analysis of the optical properties of the RuCl$_3$ nanoflakes, it is necessary to fit the measured ellipsometry angles with an appropriate multilayer model. With the aim of minimizing substrate-induced uncertainties in the determination of the refractive index of RuCl$_3$ nanoflakes, ellipsometric angles from the bare SiO$_2$/Si substrate were also measured simultaneously on spatial positions a few microns away from each RuCl$_3$ nanoflake using the same experimental conditions. The thickness of the SiO$_2$ oxide layer corresponding to each spatial location was estimated by fitting the experimental substrate data to a multilayer model using the Mueller matrix formalism, where we employed reported refractive indices for Si and SiO$_2$ \cite{henrie2004electronic}. Next, the same Mueller matrix formalism was employed to estimate the energy-dependent complex refractive index of RuCl$_3$ nanoflakes with different thicknesses. Figure \ref{fig1}(c) shows the estimated complex refractive index of RuCl$_3$ obtained from a simultaneous fitting of the $\Delta$ and $\Psi$ angles shown in Fig. \ref{fig1}(b) for different crystal thicknesses and angles of incidence. The solid lines in this figure show the results of the modelled ellipsometry angles obtained from the simultaneous fitting approach. The  ellipsometry data corresponding to RuCl$_3$ flakes with thicknesses of 2.9, 6.9, and 36.5 nm yield the same complex refractive index values, confirming the thickness-independent optical properties of this material. 

Further, we observe a very good agreement between the estimated refractive index of the nanoflakes and the optical properties reported for bulk RuCl$_3$ crystals in previous works \cite{binotto1971optical,plumb2014alpha,sandilands2016spin,sandilands2016optical}. As seen in Fig. \ref{fig1}(c), the imaginary part of the refractive index ($\kappa$) shows a clear absorption peak at $\sim$2.05 eV, in good agreement with the $\beta$ resonance reported in the imaginary part of the dielectric constant \cite{sandilands2016spin} and the real part of the optical conductivity of bulk RuCl$_3$ crystals \cite{sandilands2016optical}. The observed increase of $\kappa$ towards the edges of our measurement range also agrees well with the reported existence of stronger $\alpha$ and $\gamma$ absorption resonances at $\sim$1.2 and 3.2 eV, respectively \cite{sandilands2016spin,sandilands2016optical}.

\section{\label{sec:level3}Optical contrast analysis of R\lowercase{u}C\lowercase{l}$_3$}

Next, we employ the measured complex refractive index of RuCl$_3$ to calculate its thickness-dependent optical contrast when deposited on top of SiO$_2$/Si substrates and imaged with an optical microscope in an epi-illumination configuration. In our simulations, we adopt the following definition for the optical contrast between the nanoflake and the substrate \cite{jung2007simple,castellanos2010optical}: $OC(\lambda)=(R(\lambda)-R_0(\lambda))/(R(\lambda)+R_0(\lambda))$, where $R(\lambda)$ and $R_0(\lambda)$ represent the wavelength-dependent intensities of the light reflected by the whole heterostack (RuCl$_3$/SiO$_2$/Si) and the bare substrate (SiO$_2$/Si), respectively. To calculate $R(\lambda)$ and $R_0(\lambda)$ we employ the transfer matrix method, in which the forward- and backward-propagating plane waves form the basis into which the optical electric field in each layer is decomposed \cite{macleod2001crc}. In our simulations, we include the wavelength-dependent refractive indices of SiO$_2$ and Si \cite{henrie2004electronic}, and assume normal incidence of the illumination for simplicity. Although the angular spread of incident angles given by the numerical aperture (NA) of the illumination objective lens can play a significant role on the perceived optical contrast for thick crystals and/or 2D layers with anisotropic optical constants \cite{katzen2018rigorous,huang2019optical}, the relatively thin flakes ($<100$ nm), the low NA of the objective lens (0.42), and the measured isotropic refractive index of RuCl$_3$ justify our assumption.

\subsection{Thickness-dependent optical contrast of R\lowercase{u}C\lowercase{l}$_3$ in the \lowercase{s}RGB color space}

\begin{figure}	
    \includegraphics[scale=0.58]{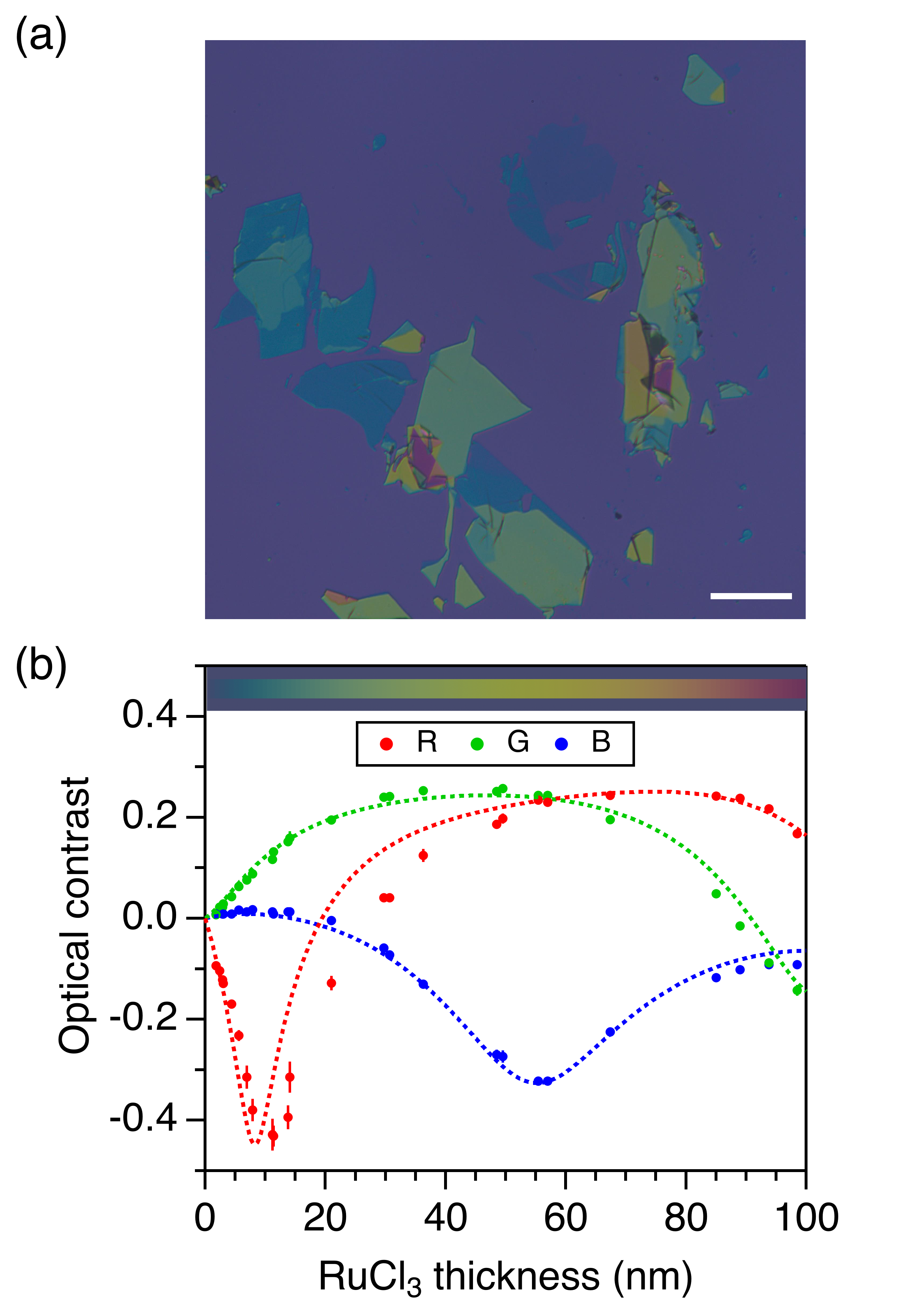}
    	
    \caption{(a) Optical microscope image, acquired in the standard RGB (sRGB) color space, of exfoliated RuCl$_3$ flakes with thicknesses ranging from $\sim$ 2 $-$ 100 nm. Scale bar 50${\mu}$m. (b) Comparison of the experimental (dots) and calculated (dashed lines) optical contrast in the sRGB color space corresponding to the red (R), green (G), and blue (B) channels. The colorscale bar shows the calculated color of RuCl$_3$ as function of flake thickness when deposited on top of a SiO$_2$/Si substrate with a SiO$_2$ thickness of 295.5 nm.}
    \label{fig2}
\end{figure}

In this section, we explore experimentally and numerically the thickness-dependent optical contrast of RuCl$_3$ on SiO$_2$/Si substrates in the standard RGB (sRGB) color space under broadband white illumination. Figure \ref{fig2}(a) shows an optical microscope image in the sRGB color space of RuCl$_3$ flakes with thicknesses in the range $\sim2-100$ nm mechanically exfoliated on a SiO$_2$/Si substrate with a SiO$_2$ thickness of 295 nm, as confirmed by spectroscopic ellipsometry. The red, green, and blue dots in Fig. \ref{fig2}(b) represent the experimental optical contrast values in the different channels (R,G, and B, respectively) extracted from the optical microscope image shown in Fig. \ref{fig2}(a) and a second  spot nearby on the same substrate. The thicknesses of the layers indicated in Fig. \ref{fig2} were estimated by a combination of AFM and spectroscopic ellipsometry measurements, while the optical contrast values and corresponding error bars represent the optical contrast value and associated uncertainty calculated from the measured $R(\lambda)$ and $R_0(\lambda)$ in each flake and a nearby spot in the substrate, respectively. The experimental values and uncertainties of $R(\lambda)$ and $R_0(\lambda)$ for all the measured flakes were obtained from the median and standard deviation, respectively, of 100 pixels.

In order to compare the experimental contrast values in the sRGB color space with the calculated optical contrast, we follow the approach described in previous works \cite{henrie2004electronic,gao2008total,muller2015visibility,spina2016rapid,chen2017layer}. In this approach, the simulated wavelength-dependent reflected intensities of the flake and bare substrate ($R(\lambda)$ and $R_0(\lambda)$) are used to calculate CIEXYZ color values, which simulate the color perceived by the human eye. Such transformation involves an integration over wavelengths including the light source spectrum (Thorlabs, Inc., product MCWHLP1) and the CIE color-matching functions to calculate the XYZ tristimulus components \cite{henrie2004electronic,gao2008total,muller2015visibility}. Finally, the calculated XYZ values are converted to the sRGB color space using a standard transformation, which takes into account the chromaticity coordinates of the sRGB color space and the reference white of the light source \cite{henrie2004electronic,Lindbloom}. 

The dashed lines in Fig. \ref{fig2}(b) show sRGB optical contrast values for the R, G, and B channels as function of flake thickness calculated following the approach described above. We note that in order to get a good agreement with the experimental values, the simulated optical contrast response of the three channels must be scaled by a normalization factor $f$ (with $f<1$). We justify the use of such normalization factor by the need to account for the reduced spatial and temporal coherence of the illumination light in our experimental setup (which we assume to be perfect in the simulations), which is known to result in a reduced visibility of the interference fringes in the reflected signals while preserving the thickness-dependent positions of the interference maxima and minima \cite{troparevsky2010transfer}. We find that a normalization factor $f=0.68$ effectively reproduces the observed optical contrast in all three RGB channels for all measured RuCl$_3$ thicknesses. Notably, the best agreement is found using the G and B channels, showing goodness-of-fit parameters (GoF) $>$ 0.98, as compared to a value of $\approx0.92$ obtained for the R channel, where $GoF=1-RMSE$, with $RMSE$ being the root-mean-square error. 

The good agreement between the experimental and simulated optical contrast allows us to calculate the apparent color of RuCl$_3$ on a SiO$_2$(295 nm)/Si substrate as a function of flake thickness. The colorbar in the top side of Fig. \ref{fig2}(b) shows the calculated apparent colors of the substrate (outer region) and the RuCl$_3$ (inner region) as a function of thickness. Overall, we find a good agreement between the measured and calculated apparent color of RuCl$_3$, which can be used as a fast method for assessing flake thicknesses, as shown previously for other different 2D materials \cite{roddaro2007optical,muller2015visibility,chen2017layer,puebla2022apparent}.

\begin{figure}
    \includegraphics[scale=0.58]{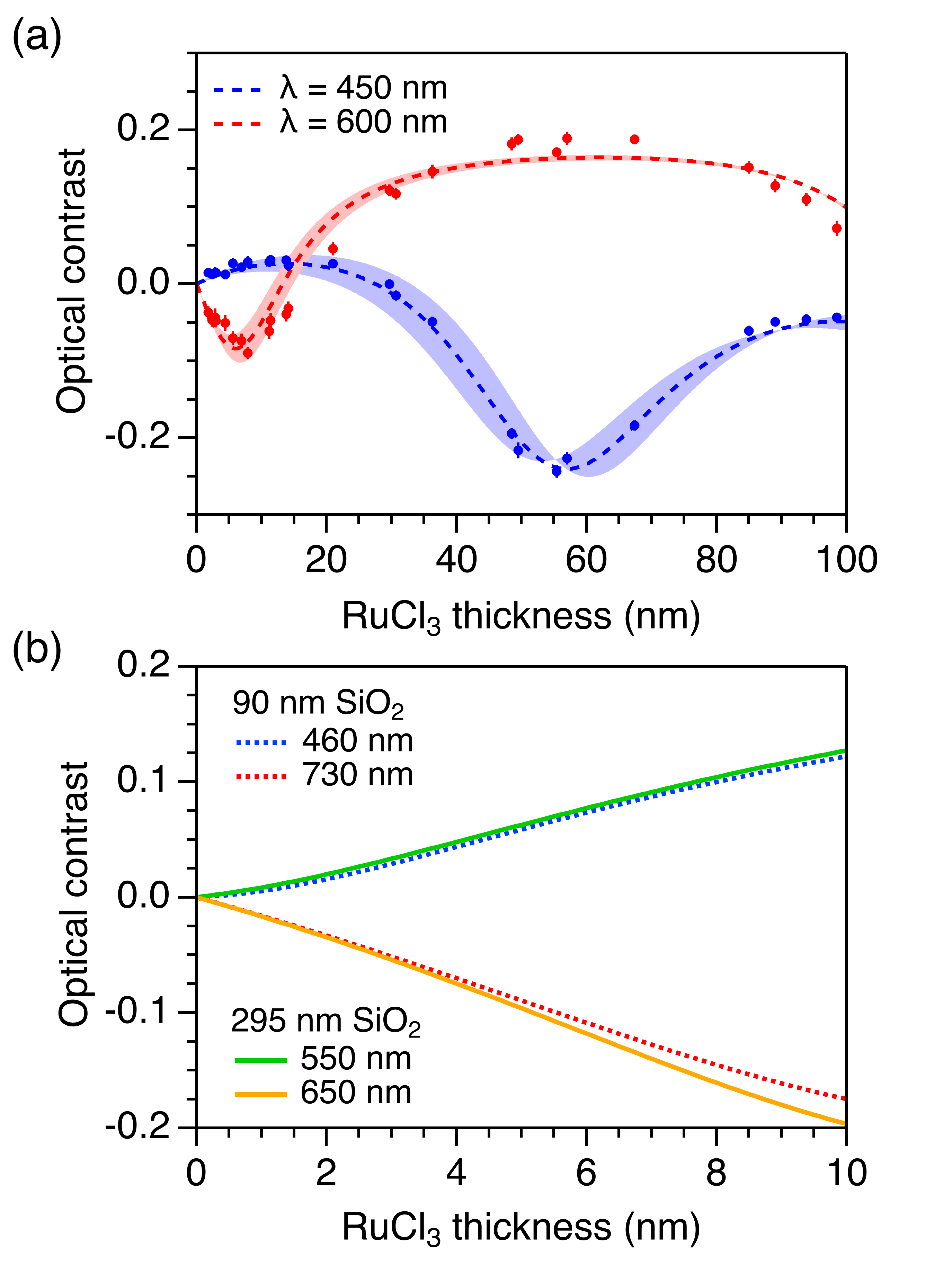}
    	
    \caption{(a) Comparison of the experimental (dots) and calculated (dashed lines) optical contrast of RuCl$_3$ of with thicknesses $\sim$ 2 $-$ 100 nm measured using bandpass filters at 450 nm (blue) and 600 nm (red) with a 10-nm bandwidth. The shaded areas represent the calculated optical contrast at the corresponding central wavelength of the filter $\pm$ 5 nm. (b) Calculated optical contrast under the monochromatic excitation wavelengths that maximise the positive and negative optical contrast of RuCl$_3$ with thicknesses in the range $\sim$ 0 $-$ 10 nm for SiO$_2$ thicknesses of 90 nm (dashed lines) 295 nm (solid lines).}
    \label{fig3}
\end{figure}

\subsection{Thickness-dependent optical contrast of R\lowercase{u}C\lowercase{l}$_3$ under narrow-band illumination}

Next, we explore the effects that the inclusion of narrow optical bandpass filters in the illumination path has on the optical contrast characterisation of RuCl$_3$. This approach has previously shown to reproduce to a good extent the optical contrast simulated under monochromatic illumination in other 2D crystals \cite{blake2007making,castellanos2010optical,brotons2017optical,krevcmarova2019optical}. The blue and red dots in Fig. \ref{fig3}(a) represent the experimental contrast values measured for the same RuCl$_3$ flakes shown in Fig. \ref{fig2} when filtering the illumination spectrum with 10-nm-bandwidth filters centred at 450 and 600 nm, respectively. The blue and red dashed lines indicate the simulated thickness-dependent evolution of the optical contrast assuming monochromatic excitation at 450 and 600 nm, respectively. The blue and red shaded areas show the simulated optical contrast for monochromatic illumination in the range $450\pm5$ and $600\pm5$ nm, respectively. Similar to the results under broadband illumination, we observe that the simulated optical contrast agrees well with the experimental values across the whole thickness range. Again, we find a better agreement between simulated and experimental values for the shorter illumination wavelength, with GoF values of 0.99 and 0.98 for the 450 nm and 600 nm filters, respectively. 

Once more, the good agreement between the optical contrast measured with bandpass filters and the simulated optical contrast under monochromatic illumination allows us to explore the monochromatic wavelengths that enhance the optical contrast response of RuCl$_3$ with thicknesses below 10 nm on top of Si substrates with thermally grown SiO$_2$ layers. Note that an enhanced optical contrast response arises from a combination of large absolute optical contrast with the underlying substrate and a linear dependence with RuCl$_3$ thickness. Figure \ref{fig3}(b) summarises the results of our analysis for two standard SiO$_2$ thicknesses of 90 and 295 nm. Our results suggest that monochromatic illumination wavelengths of $\sim730$ nm and $\sim650$ nm yield the largest optical contrast values for 90-nm and 295-nm-thick SiO$_2$ layers, respectively, while showing an almost linear dependence with RuCl$_3$ thickness in the explored range. We note that in both cases the optimised monochromatic illumination wavelengths result in a negative optical contrast, i.e., the RuCl$_3$ crystal appears darker than the substrate. We find that monochromatic illumination wavelengths of $\sim460$ nm and $\sim550$ nm give rise to the opposite case (i.e., RuCl$_3$ appears brighter than the substrate) for 90-nm and 295-nm-thick SiO$_2$ layers, respectively, at the expense of a slightly lower absolute optical contrast.

\subsection{R\lowercase{u}C\lowercase{l}$_3$ thickness estimation from optical contrast measurements}

Finally, in this section we benchmark the performance of the optical contrast technique to estimate the thickness of RuCl$_3$ nanoflakes using our calibrated optical microscope and experimental optical contrast dataset shown in Fig. \ref{fig2} and Fig. \ref{fig3}. In order to do so, we employ an algorithm that estimates the value of the RuCl$_3$ thickness ($d$) that minimizes the following quantity: 
\begin{align}
    RMSE^{total}(d)=\sqrt{\sum_{i}^N\frac{[OC_{exp}^i-OC_{sim}^i(d)]^2}{N}},
\end{align}
\label{minimization}
with $N$ representing the total number of experimental and simulated optical contrast illumination/detection modes $i$ considered for the minimization algorithm (i.e., sRGB detection for white illumination or bandpass illumination). In our algorithm, we follow a two-step approach. We start by considering only the two optical contrast configurations that show the best overall agreement between experimental and simulated values (G and B channel detection under broadband illumination) and estimate the crystal thickness ($d$) that minimizes Eq. (1). In the second step, if the value of $d$ returned by the initial minimization of Eq. (1) for the G and B channels falls outside the range $3.5 \leq d \leq55$ nm (i.e. where the R channel shows a good agreement between experiment and simulations), we minimize again Eq. (1) with the addition of the R channel. This two-step minimization procedure results in an estimated thickness with a mean deviation $\overline{\Delta d}=-0.2$ nm from the crystal thickness estimated by AFM/ellipsometry in the range $d<100$ nm with a standard deviation of 1.2 nm. 

\begin{figure}	
    \includegraphics[scale=0.54]{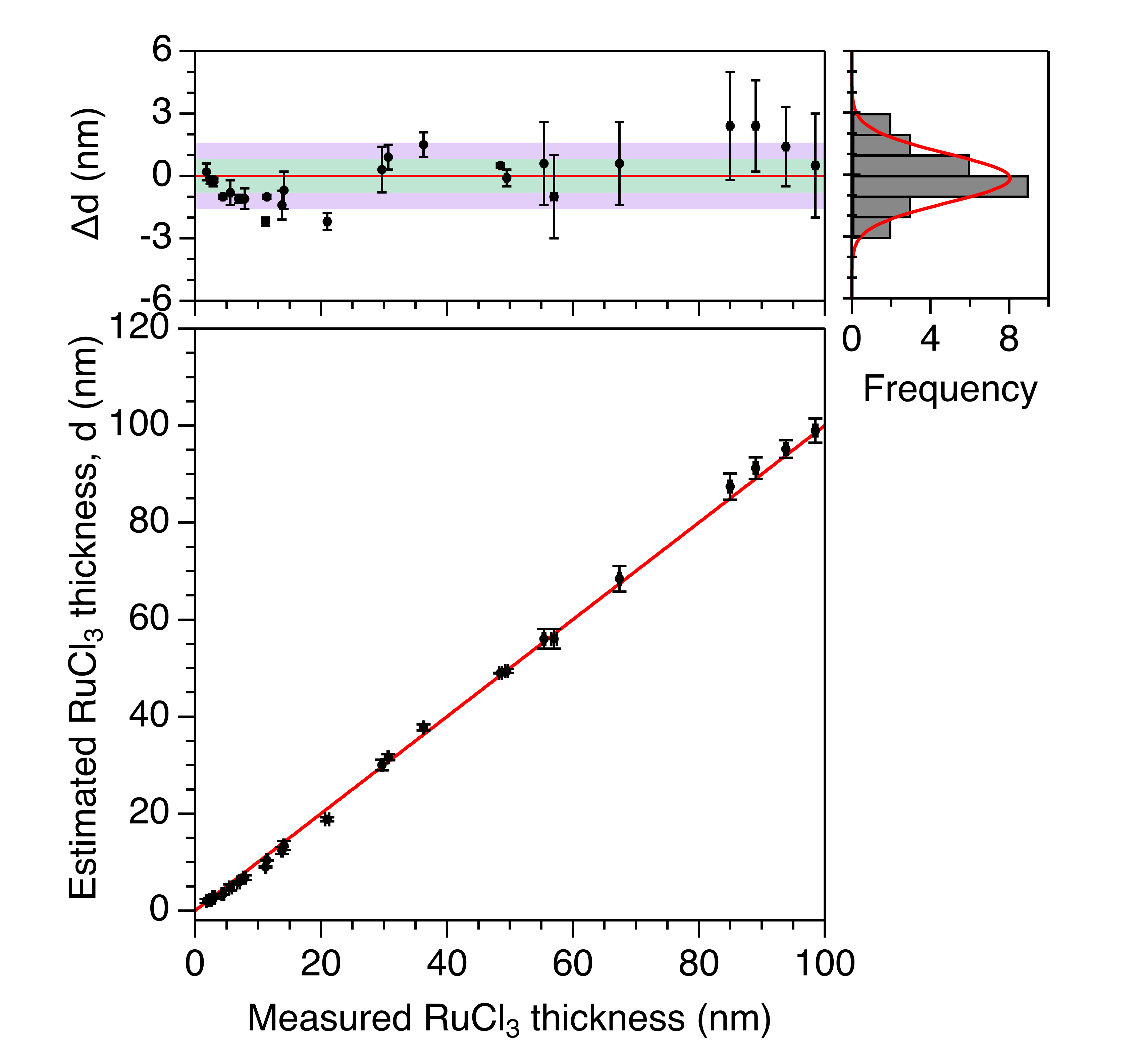}
    	
    \caption{Bottom panel: RuCl$_3$ thickness estimated from the experimental optical contrast obtained by a combination of sRGB and monochromatic detection as a function of the thickness determined by independent AFM and/or spectroscopic ellipsometry measurements. The horizontal error bars represent the uncertainty associated with the independent thickness estimation. The vertical error bars represent the uncertainty resulting from our thickness estimation approach, which we define as the thickness range for which the $RMSE^{total}$ defined by Eq. (1) is less than or equal to two times the $RMSE^{total}$ corresponding to the estimated thickness. The red line represents the ideal case, where the thickness inferred from the optical contrast equals the measured thickness. Top panel: difference between the estimated thickness inferred from the optical contrast and the measured RuCl$_3$ thickness ($\Delta d$). The green and purple shaded areas represent $\Delta d$ intervals of $\pm$1 and $\pm$2 RuCl$_3$ monolayers, respectively. The histogram in the top-right panel identifies the statistics using monolayer thickness binning.}
    \label{fig4}
\end{figure}

Finally, we note that the inclusion of an additional measurement mode (450 nm band-pass detection) in the second stage of the two-step minimization algorithm leads to a slight improvement in the thickness estimation performance. The bottom panel of Fig. \ref{fig4} shows the RuCl$_3$ thickness estimated from the experimental optical contrast as a function of
the thickness determined by AFM and spectroscopic ellipsometry measurements. The red line represents the ideal case, where the thickness inferred from the optical contrast equals the measured thickness. The top panel shows the difference between the estimated thickness inferred from the optical contrast and the measured RuCl$_3$ thickness ($\Delta d$), with the green and purple shaded areas representing $\Delta d$ intervals of $\pm$1 and $\pm$2 RuCl$_3$ monolayers, respectively. Finally, the top right panel shows a histogram of the corresponding $\Delta d$ values, which yields a $\overline{\Delta d}=-0.2$ nm for the whole thickness range with a standard deviation of 1 nm, i.e., close to the thickness of a single RuCl$_3$ monolayer (0.8 nm \cite{zhou2019possible,lee2021multiple}).

\section{Conclusions}
In summary, we report the complex refractive index of $\alpha$-RuCl$_3$ nanoflakes with thicknesses below 40 nm in the visible and near-infrared wavelength range. Our results show that the optical properties of the nanoflakes are independent of the crystal thickness and agree well with the optical constants reported for bulk samples \cite{binotto1971optical,plumb2014alpha,sandilands2016spin,sandilands2016optical}. Compared to other 2D crystals such as transition-metal dichalcogenide semiconductors \cite{hsu2019thickness}, group III–VI metal chalcogenide semiconductors such as InSe \cite{brotons2016nanotexturing}, or hybrid perovskites \cite{song2020determination}, which show a strong modulation of their dielectric response with the thickness for few-layer crystals, the thickness-independent optical response of RuCl$_3$ enables a straightforward and reliable comparison between the experimental and simulated optical reflectance contrast as function of crystal thickness. In this context, we show that transfer-matrix based simulations of the thickness-dependent optical contrast of RuCl$_3$ on oxidized silicon substrates reproduce with good agreement the experimental optical contrast obtained under different illumination/detection configurations, including sRGB color-space detection for broadband and narrow-band illumination. Finally, we show that optical contrast imaging allows us to retrieve the thickness of the RuCl$_3$ nanoflakes exfoliated on an oxidized silicon substrate with a mean deviation of $-0.2$ nm for thicknesses below 100 nm with a standard deviation of only 1 nm across the whole thickness range. These results demonstrate the potential of optical contrast analysis as a non-invasive, fast, and reliable technique to estimate the thickness of RuCl$_3$ nanoflakes.

\begin{acknowledgments}This work was supported by the EPSRC (grant nos. EP/P029892/1 and EP/L015110/1), and the ERC (grant no. 725920) (D.A.-P., S. O., and B.D.G). The work of Y.W. and K.S.B. was supported by the National Science Foundation (NSF) EPMD program via grant EPMD-2211334. Crystal growth at ORNL was supported by the U.S. Department of Energy, Office of Science, National Quantum Information Science Research Centers, Quantum Science Center. M.B.-G. is supported by a Royal Society University Research Fellowship. B.D.G. is supported by a Chair in Emerging Technology from the Royal Academy of Engineering.
\end{acknowledgments}

\section*{Author declarations}
\subsection*{Conflict of Interest}
The authors have no conflicts to disclose.

\section*{Data Availability}
The dataset generated and analyzed during the current study is available at: https://researchportal.hw.ac.uk/en/persons/mauro-brotons-i-gisbert/datasets/

\bibliography{Bibliography}

\end{document}